# Long-lived spin coherence in silicon with an electrical spin trap readout

G. W. Morley,[1,*] D. R. McCamey,[2] H. A. Seipel,[2] L.-C. Brunel,[3] J. van Tol,[3] and C. Boehme[2,†]

[1]*London Centre for Nanotechnology and Department of Physics and Astronomy, University College London, London WC1H 0AH, UK*

[2]*Department of Physics, University of Utah, 115 South 1400 East Rm 201, Salt Lake City, Utah 84112, USA*

[3]*National High Magnetic Field Laboratory at Florida State University, Tallahassee Florida 32310, USA*

Pulsed electrically-detected magnetic resonance of phosphorous ($^{31}$P) in bulk crystalline silicon at very high magnetic fields ($B_0 > 8.5$ T) and low temperatures ($T = 2.8$ K) is presented. We find that the spin-dependent capture and reemission of highly polarized (>95%) conduction electrons by equally highly polarized $^{31}$P donor electrons introduces less decoherence than other mechanisms for spin-to-charge conversion. This allows the electrical detection of spin coherence times in excess of 100 μs: 50 times longer than the previous maximum for electrically-detected spin readout experiments.

Electrical (current) detection of pulsed magnetic resonance has previously been used to manipulate and readout the state of a single electron spin in GaAs quantum dots [1], but the nuclear spins in this material provide severe decoherence [2]. Silicon has fewer nuclear spins than GaAs, and smaller spin-orbit coupling, so the decoherence times of electron spins in silicon are more favourable by around three orders of magnitude: phosphorus donors in silicon (Si:P) are the only electron spins that have been shown to store quantum information for times in excess of 10 ms [3], supporting proposals for using them as qubits [4,5].

These quantum computing proposals require the ability to readout the state of a single spin, and electrical detection is the leading candidate for this [6]. Reading out electron spin information using electron spin resonance (ESR) with conventional microwave detection requires a sample of $>10^8$ electrons [3]. Optical readout is an alternative [7], but electrically-detected magnetic resonance (EDMR) has provided the highest sensitivity so far for silicon samples: one group detected a single unidentified electron spin [8], and another studied fewer than 100 phosphorus donors [6].

Electrically-detected spin coherence has been demonstrated in silicon using pulsed (p) EDMR of dangling bonds (db) [9] and phosphorus-db pairs [10,11]. The coherence of conduction electron spins in pure silicon has also recently been studied electrically [12]. However, as with the pEDMR of GaAs and all other samples, the electrically-detected spin coherence observed so far in silicon has survived for less than 2 μs, removing the key advantage of silicon for quantum computation.

In the following, we present pEDMR experiments conducted at magnetic fields over 25 times higher than any previous pEDMR experiment. At the low field of ~0.33 T used to date for pEDMR, spin coherence in silicon has been limited to 2 μs due to the fast spin-dependent recombination of photoelectron-hole pairs [10] which allows the electric readout. We show here that, at 8.6 T, one can use the spin-trap mechanism proposed by Thornton and Honig [13] in order to detect the $^{31}$P state electrically. This mechanism does not require the presence of coherence limiting probe spins (in contrast to spin-dependent recombination based schemes) and quenches spin coherence times solely by charge carrier trapping/reemmission.

EDMR requires that changing a spin state alters the conductivity, and several microscopic mechanisms can be responsible for this in Si:P samples [6,9,10,13,14,15,16]. As in most previous EDMR experiments on Si:P, we obtained bulk conductivity by illuminating with photons of greater energy than the silicon band gap to create electron-hole pairs.

Our experiments were performed at 240 GHz with the quasi-optical ESR spectrometer [17] that has been developed at the National High Magnetic Field Laboratory in Tallahassee, Florida. This spectrometer has recently been upgraded [18] to operate in pulsed mode and at frequencies up to 336 GHz. Pulsed ESR at magnetic fields above 3 T is complicated by the lack of commercially-available high-power radiation sources at frequencies of 100-600 GHz. The sample must experience high powers so that spin manipulations take place before decoherence occurs, and we achieve this with a Fabry-Pérot resonator which concentrates the mm-wave power at the sample.

Our sample [19] uses five interdigitated Al wires which overlap for 1 mm, as shown schematically in the inset of Fig. 1(a). The wires are 100 nm thick, 10 μm wide and are separated by 10 μm. The contacts were made with optical lithography on a 330 μm thick (111) oriented prime grade Cz-grown crystalline silicon wafer containing [P] $\approx 10^{15}$ cm$^{-3}$ phosphorus dopants, after an HF etch dip was executed to remove the native oxide. For all of the data shown here, the sample temperature was 2.8 K and continuous irradiation with a filtered Xe lamp generated a photocurrent $I$ = 60 nA.

EDMR spectra of Si:P at high magnetic fields of 3.5-7.1 T and temperatures of 1.4-5 K have been attributed to spin-dependent trapping of photoexcited conduction electrons [13,20]. Fig. 1(a) reproduces these spectra at the higher magnetic field of $B_0 \sim$ 8.6 T. The current is reduced when spin resonance conditions are satisfied. For higher temperatures this mechanism becomes less significant as the thermal energy becomes larger than the trapping energy [13]. Spin dependent recombination of electron-hole pairs is not significant for a magnetic field of 8.6 T and a temperature of 2.8 K because the large electronic polarization favours spin-dependent trapping. The high-field resonance in Fig. 1(a) is larger than the low-field resonance due to the anti-polarization of $^{31}$P nuclear spins [21]. Sample heating due to the resonant ESR absorption of energy is negligible for these low P concentrations so the signal we record cannot be explained as bolometry. In Fig. 1(a) the phosphorus spins produce two large resonance peaks centred near $B_0 = 8.58$ T, whilst the broader silicon db signal is barely visible near $B_0 = 8.57$ T, demonstrating that the $^{31}$P-db mechanism usually seen at low fields [10,11] is insignificant.

Figs. 1(b)-1(c3) depict the EDMR mechanism [13] that is confirmed in this study whereby Pauli exclusion prevents conduction electron trapping unless millimetre-waves flip the spin of the phosphorus donor electron. $D^0$ is the neutral donor state in which one electron is bound to the phosphorus nucleus. Trapping a further electron produces the $D^-$ state which has a higher energy that is still below the conduction band [22].

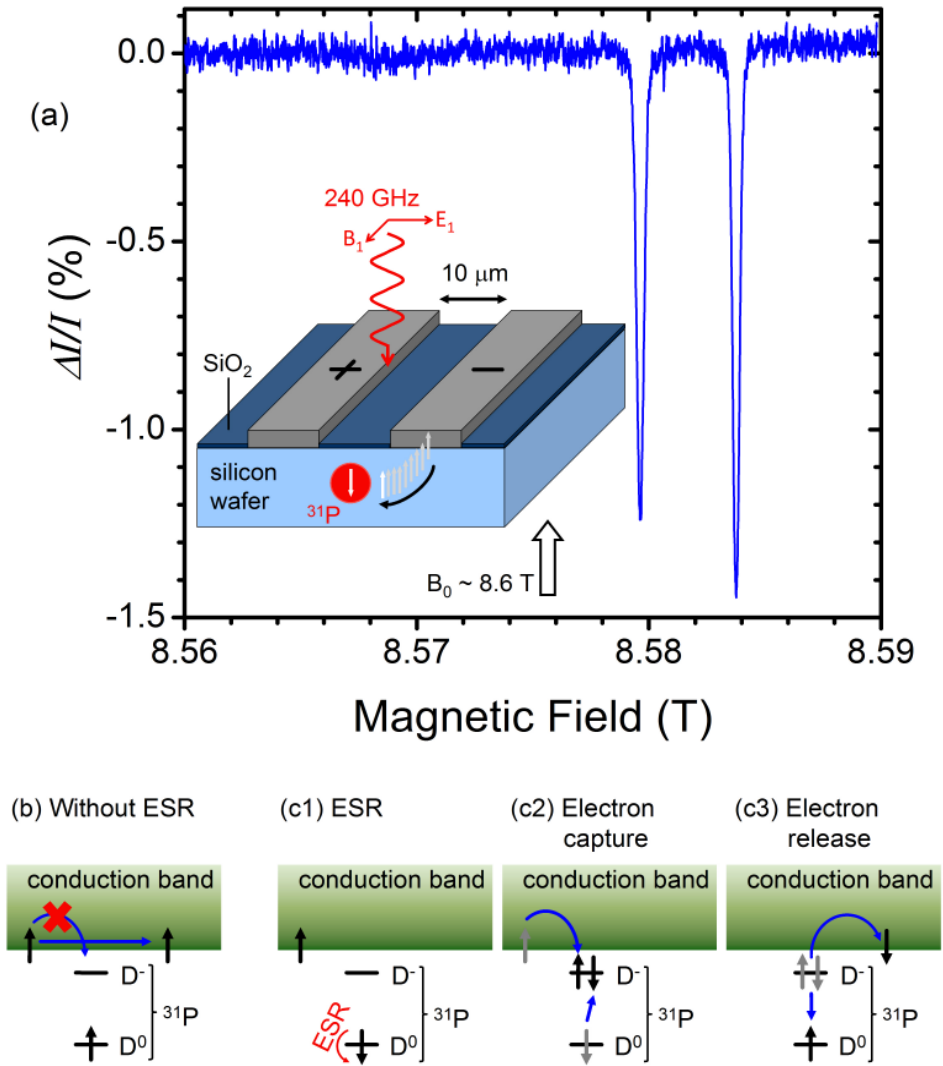

**FIG. 1. (color online) (a) 240 GHz electrically-detected magnetic resonance spectrum of phosphorus donors in crystalline silicon. The two large resonances are separated by 4.2 mT: a signature of the hyperfine interaction with a phosphorus nuclear spin. The inset is a schematic of our experiment, shown in more detail in Fig. 1 of Ref. 19. (b) The spin of an excess electron in the conduction band is initially aligned with the applied field so the Pauli exclusion principle means it avoids the co-aligned phosphorus spin. (c1) ESR flips the phosphorus spin. (c2) The conduction electron is trapped, reducing the current. (c3) An electron is reemitted, leaving the electron that remains in the opposite spin state.**

This trapping/reemission effect increases with electronic polarization, explaining why it has not been reported for magnetic fields below 3.5 T. The electronic polarization (ratio of net polarized spins to total number of spins) used here was calculated to be >95%. This would provide a suitable starting state for electron spin-based quantum computation, even with many qubits.

To understand the dynamics of the process responsible for the strong $^{31}$P signal at $T$ = 2.8 K, we have carried out pulsed (p) EDMR measurements similar to the recent coherent spin manipulation of P-db spins at low magnetic fields [10,11]. With these experiments, pulsed magnetic resonance is used to coherently control spin dynamics while current measurements probe the spin motion.

Fig. 2 shows the transient electrical response to a single mm-wave pulse. With the magnetic field set to be away from the phosphorus resonance there is no signal (not shown). This is in marked contrast to the microwave artifacts that are observed at lower fields [9]. The sensitivity of some previous low-field experiments has been limited by these microwave artifacts, indicating that greater sensitivity may be available with pEDMR at high fields and low temperatures. The number of spins in our experiment is ~$10^9$, and for the current transient in Fig. 2 the single-shot sensitivity is ~$5 \times 10^7$ spins: around 100 times higher than with mm-wave detection [18]. Furthermore, it has been found that the signal to noise ratio is independent of sample size in EDMR [6], so if our experiment scales in the same way, single electron spin detection (and therefore single nuclear spin readout [23]) could be reached with a (100 nm)$^3$ silicon sample.

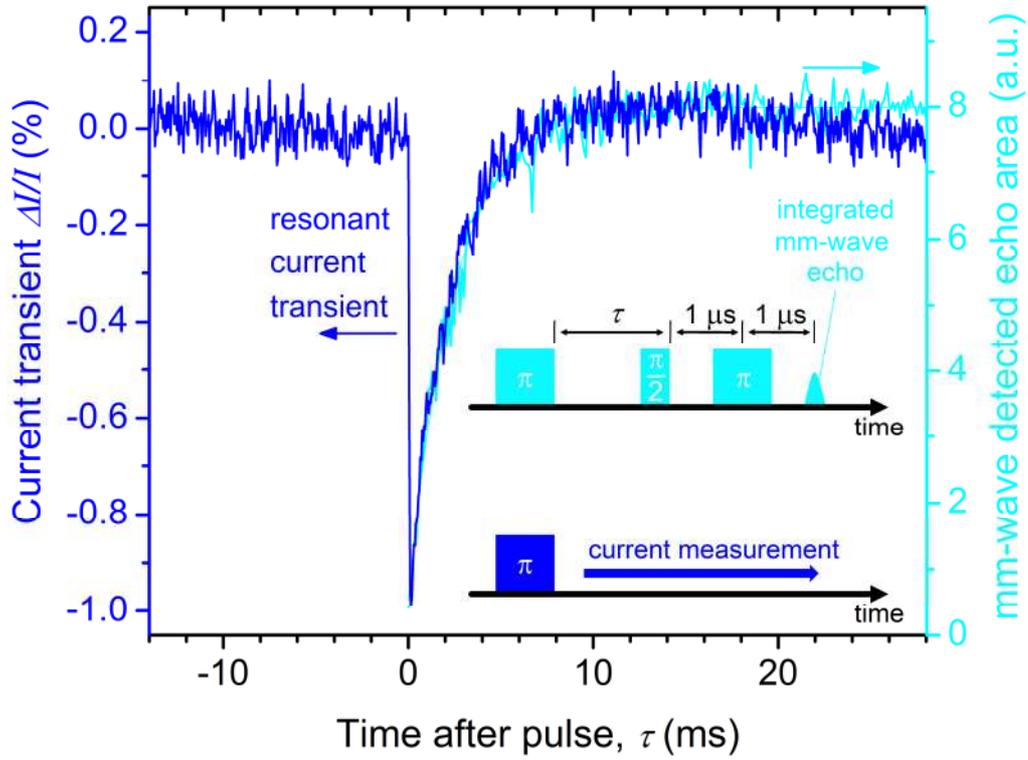

**FIG. 2. (color online) Comparison of electrical and mm-wave ESR detected spin readouts: the plots show an ESR induced current response (dark data) and the ESR-measured inversion recovery (lighter data). In these experiments a short (480 ns) pulse of 240 GHz radiation excites the silicon sample before detection in either the mm-wave response or the electrical current. On resonance, the current is strongly reduced, exponentially returning towards equilibrium with a decay time of 2.5 ±0.1 ms. The mm-wave detected spin-lattice ($T_1$) relaxation time measured with the 'inversion recovery' pulse sequence [24] is also 2.5 ±0.1 ms. Both data sets exhibit the same decay behaviour.**

With the magnetic field set to the phosphorus resonance the current is seen to drop within 100 μs, and then recover with an exponential time constant of 2.5 ms. In the trapping/reemission picture described above, we attribute this recovery to the gradual release of trapped electrons by the phosphorus dopants. This results, for each trapping/reemission, in the formation of an anticorrelated spin pair consisting of the donor electron and the emitted (flying) electron. Note that the spins within this electron pair are anticorrelated, and either $|\uparrow\downarrow\rangle$ or $|\downarrow\uparrow\rangle$ is produced randomly. As the process of trapping and reemission randomizes the spin state of the electron that remains on the phosphorus dopant, this constitutes a $T_1$ relaxation process.

The dynamics of spin $T_1$ processes can be measured with pulsed ESR experiments, so a test of the trapping/reemission picture becomes possible: the measured current transient and the pulsed ESR-measured $T_1$ time must agree if this model is correct. Pulsed ESR uses inversion recovery experiments to measure $T_1$: the return to polarization of antipolarized spins is recorded with the pulse sequence $\pi - \tau - \pi/2 - t - \pi - t -$ echo [24]. In this sequence, the first pulse inverts the spins, and the rest of the sequence measures the magnetization returning to equilibrium as a function of the delay time $\tau$. For the Si:P inversion recovery data shown in Fig. 2 the delay time $t = 1$ µs, $\pi = 600$ ns, and the delay time $\tau$ is incremented on the $x$-axis. The comparison with the real-time pEDMR current transient shows an excellent agreement, confirming the trapping/reemission model.

The data in Fig. 2 show that the spin-dependent trapping/reemission process provides a $^{31}$P donor electron spin readout with two timescales: 100 µs and 2.5 ms. As shown below, one can attribute these times to the processes of trapping and reemission respectively. Both of these timescales are significantly longer than the previous barrier of 2 µs which was found for the spin-dependent $^{31}$P-db electron-hole recombination mechanism [10]. We want to stress however, that as all EDMR experiments require electronic processes for spin readout, these will always introduce barriers to coherence. For the case of the trapping/reemission process discussed here, the barrier is established by the trapping (or capture) time of a free charge carrier and as such, it can be controlled by the excess charge carrier density within the sample.

In order to investigate the influence of the trapping process on spin relaxation, we studied spin coherence using the trapping/reemission mechanism. First, we carried out electrically-detected transient nutation experiments (not shown) which revealed Rabi oscillations that decay after a few µs. These measurements provided a measure for the pulse duration needed to produce a spin rotation of 180° (480 ns), and this was used as a π pulse for the experiments presented in Figs. 2 and 3. The decay of the Rabi-oscillations is only a lower limit for the decoherence time.

To measure the quantum coherence time ($T_2$) of the system we carried out electrically-detected spin-echo experiments [10]. The three-pulse sequence used for this is shown schematically in the inset of Fig. 3(a). The experiment begins with a conventional Hahn echo sequence consisting of a π/2 pulse separated by a time $\tau$ from a π pulse; this produces a magnetisation perpendicular to the 8.6 T applied magnetic field. An extra π/2 readout pulse is then swept through the Hahn echo sequence to produce the data in Fig. 3(a). The readout pulse is needed because the measured current is sensitive to the magnetisation parallel to the applied magnetic field: the difference between spin up and spin down. The 'V' shape data observed in the leftmost pulse is due to the difference between a π pulse (beginning and end of 'V') and a π/2 pulse (centre of 'V'). The echo increases the current because it leaves the P spin aligned with the applied field.

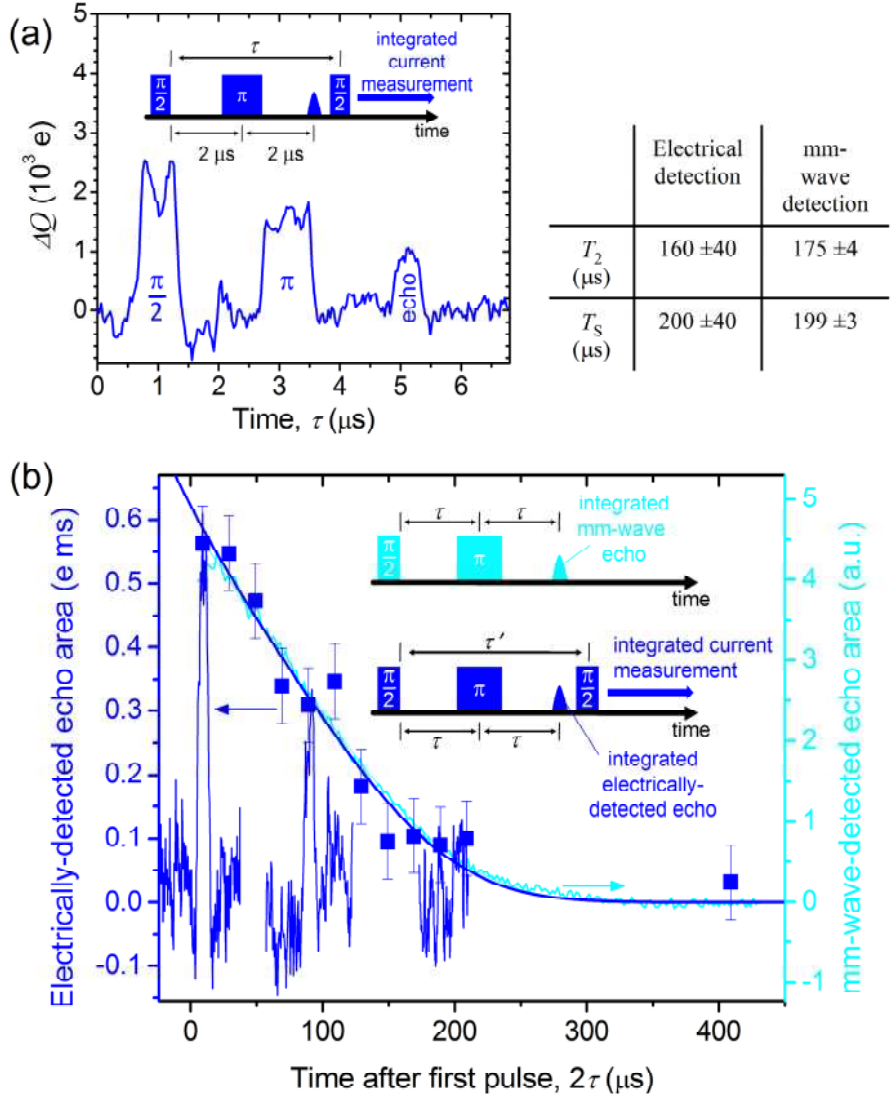

**FIG. 3. (color online) (a) An electrically-detected spin echo observed using the pulse sequence illustrated in the inset. (b) Comparison of electrically-detected (dark data) and radiation-detected (lighter data) spin echo decays which reveals the same value for $T_2$. The smooth decay curves are fits with the function exp(-$2\tau/T_2$-$8\tau^3/T_S^3$) which has previously [3,25] been used to describe Si:P with naturally occurring $^{29}$Si. $T_2$ is the intrinsic spin coherence time and the dephasing due to $^{29}$Si nuclear spins is characterized by $T_S$. Three electrically-detected echoes are shown with their time axes expanded by a factor of twenty for clarity, and with their y-axes units equal (but arbitrary).**

The square points in Fig. 3(b) reveal the coherence decay ($T_2$) times of the electrically-detected spin echoes. The dependence of echo intensity on the pulse separation time $\tau$ reveals $T_2$ = 160 µs. This time was obtained by fitting with the function $\exp(-2\tau/T_2 - 8\tau^3/T_S^3)$, which has been found [3,25] to best describe samples of Si:P containing a natural (4.7%) concentration of $^{29}$Si. This value of $T_2$ can be compared with the values previously measured using traditional microwave detection of 240 µs [26], 300 µs [25] and 2800 µs [3]. Previous electrically-detected measurement of Si:P at the lower field of 0.35 T found $T_2$ < 2 µs [10, 27]. Whilst these experiments used higher temperatures (up to 6.5 K) and P concentrations (up to $10^{17}$ cm$^{-3}$), the $T_2$ was nonetheless limited by the fixed electronic transition rate of the low-field EDMR mechanism. We have been able to record a longer coherence time by using a different mechanism, with a much slower transition rate, which may be varied by changing the conduction electron density. The decoherence in our experiment due to $^{29}$Si nuclear spins gives $T_S$ = 200 µs. Performing the fit by setting $T_S = \infty$ provides a simple exponential which is not such a good fit; the combined dephasing time then is $T_2^{\text{exponential}}$ = 108 ±14 µs.

To confirm that the electrically-measured $T_2$ times are the true coherence times of the Si:P spin, we also measure Hahn echo decays under identical experimental conditions, detecting the mm-wave emission. Fig. 3(b) and the inset table show that the $T_2$ values obtained from the electrically-detected three-pulse sequence and the radiation-detected two-pulse sequence fully agree. The mm-wave detected data are clearly non-exponential, being well fit by the function $\exp(-2\tau/T_2 - 8\tau^3/T_S^3)$ used above.

As the $T_2$ time (from Fig. 3(b)) and the drop time of the electrically-detected transient (from Fig. 2) are both around 100 µs, we attribute the $T_2$ decay to the trapping of conduction electrons by the phosphorus dopants. The trapping/reemission process works through rapid electron capture ($\approx$ 100 µs) as soon as the spin state of the phosphorous matches the opposite polarization of the excess electron ensemble and a slow reemission (≈ 2.5 ms) possibly due to thermal excitation. It may be possible to manipulate both the $T_2$ by changing the conduction electron density (e.g. via the light intensity) and also the emission time through temperature or bias voltage.

In conclusion, we have investigated the properties of a $^{31}$P spin-trap mechanism as an electrical readout for $^{31}$P electron and nuclear spins in silicon that detects coherent spin motion without compromising the intrinsic coherence times of the $^{31}$P states. We anticipate that this mechanism will allow single spin detection with appropriate sample geometry. Our investigation of the qualitative and quantitative dynamics of this mechanism shows that it permits spin coherence times longer than 100 μs. This demonstrates that electrically-detected phosphorus spin states in silicon work as even better qubits at the high magnetic fields which can initialize them.

This work was supported by Visiting Scientist Program Grant 7300-100 from the National High Magnetic Field Laboratory (NHMFL). The NHMFL is supported by NSF Cooperative Agreement No. DMR-0084173, by the State of Florida, and by the DOE. GWM was supported by the EPSRC through grants GR/S23506, EP/D049717/1 and EP/F04139X/1.

* g.morley@ucl.ac.uk

† boehme@physics.utah.edu